\documentclass[sigconf, nonacm=true]{acmart}
\setcopyright{none}
\AtBeginDocument{%
  }

\copyrightyear{2026}
\acmYear{2026}
\setcopyright{cc}
\setcctype{by}
\acmConference[]{}{}{}
\acmBooktitle{}\acmDOI{}
\acmISBN{}

\newif\ifdraft
\draftfalse
\usepackage[english]{babel}
\usepackage{xcolor}
\usepackage{xspace}
\usepackage{booktabs}
\usepackage{tabularx}

\usepackage{subcaption}
\usepackage[inline]{enumitem}
\usepackage{todonotes}
\usepackage{makecell}
\usepackage{siunitx}
\usepackage{multirow}
\usepackage{xparse}
\usepackage{pifont}         %
\usepackage{fontawesome5}
\usepackage{array}
\usepackage{dcolumn}
\usepackage{tcolorbox}

\usepackage[utf8]{inputenc} %

\usepackage{xparse}
\usepackage{url} % or hyperref

\NewDocumentCommand{\urldate}{m o}{%
  \url{#1}%
  \space(accessed: %
  \IfValueTF{#2}{#2}{March 2nd, 2026})%
}

\usepackage{svg}

\newcommand{\iconAca}{{\includegraphics[height=1em]{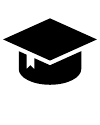}}}
\newcommand{\iconInd}{{\includegraphics[height=1em]{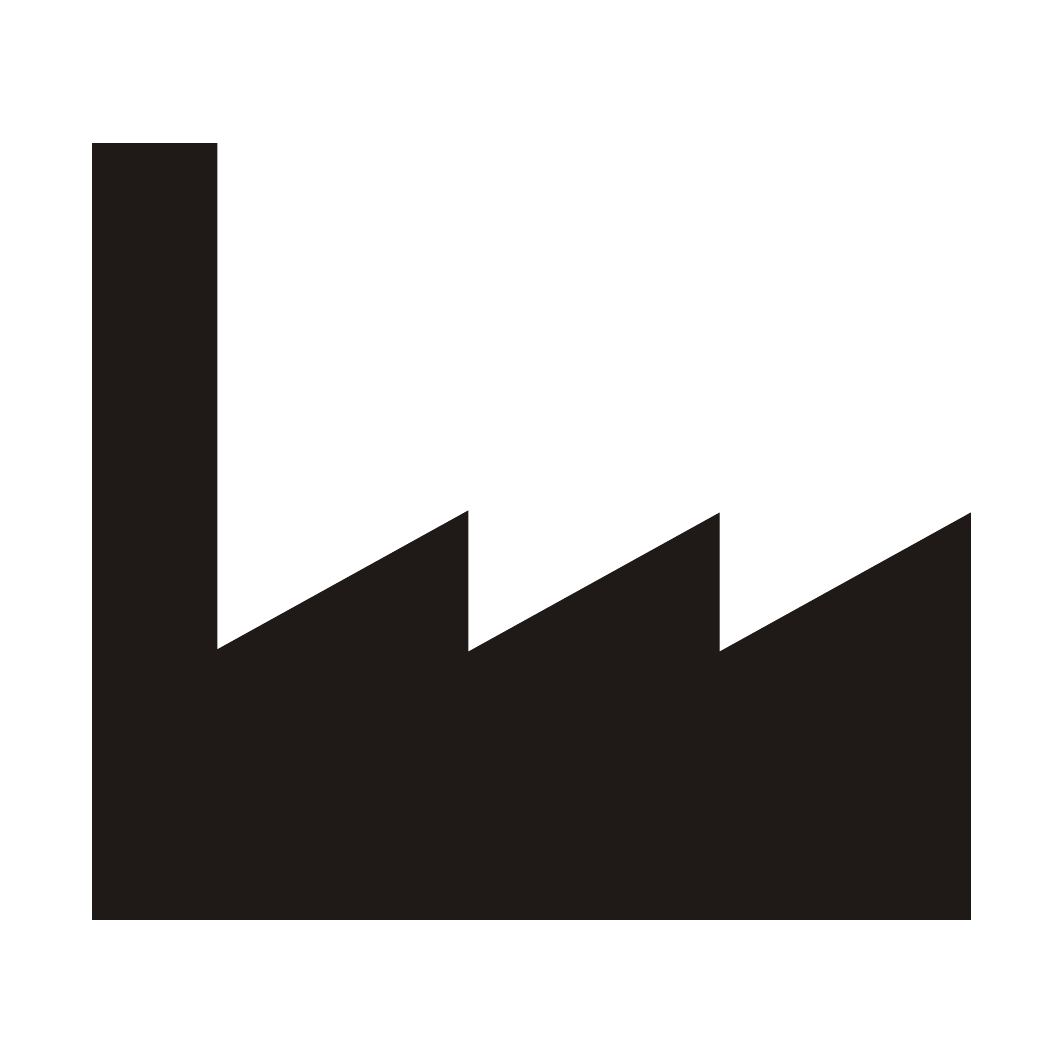}}}

\newcommand{\iconONG}{{\includegraphics[height=1em]{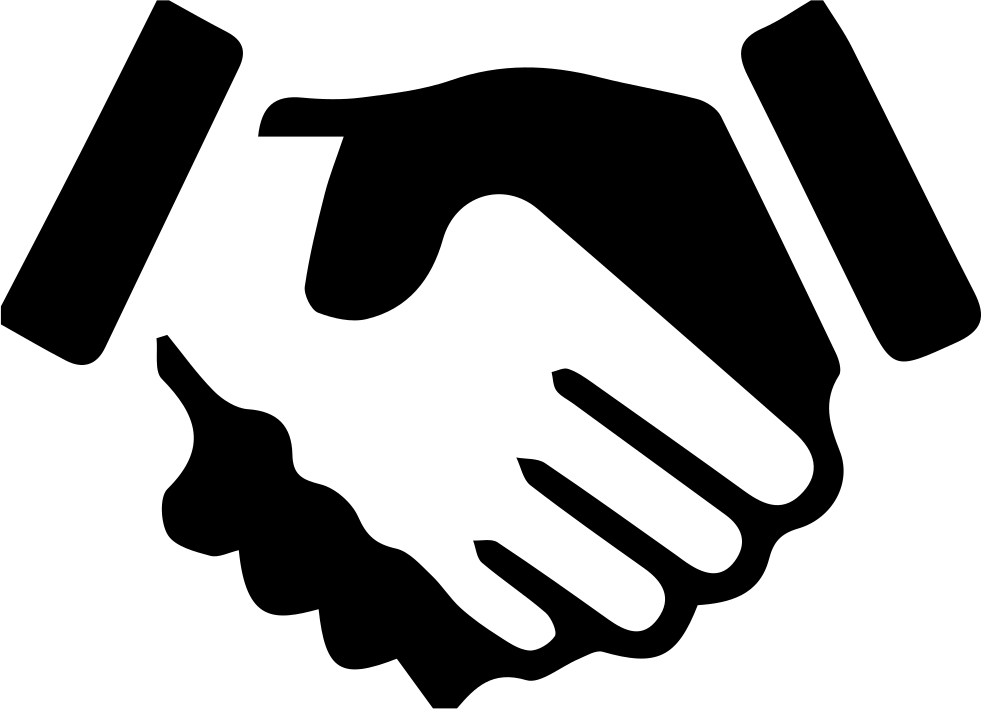}}}

\newcommand{\customSubsectionWithDate}[2]{%
  \vspace{10pt}
  \noindent
  \emph{#1 (#2).}}

\newcommand{\ansRQ}[2]{ \paragraph{\textbf{#1 (ANS-RQ#2)}}}

\NewDocumentCommand{\stput}{o}{%
  \IfNoValueTF{#1}
    {sample throughput\xspace}%
    {Sample throughput\xspace}%
}

\NewDocumentCommand{\ntput}{o}{%
  \IfNoValueTF{#1}
    {network throughput\xspace}%
    {Network throughput\xspace}%
}

\ifdraft
	\newcommand{\cg}[1]{\todo[inline,author=CG]{#1}}
	\newcommand{\ap}[1]{\todo[inline,author=AP]{#1}}
	\newcommand{\ga}[1]{\todo[inline,author=GA]{#1}}
    \newcommand{\vp}[1]{\todo[inline,author=VP]{#1}}

	\specialcomment{rewrite}{\begingroup\color{red}}{\endgroup}
	\specialcomment{draft}{\begingroup\color{blue}}{\endgroup}
\else
	\newcommand{\vp}[1]{}
	\newcommand{\ga}[1]{}
	\newcommand{\cg}[1]{}
	\newcommand{\ap}[1]{}
	
	\specialcomment{rewrite}{\begingroup}{\endgroup}
        \excludecomment{rewrite}
	\specialcomment{draft}{\begingroup}{\endgroup}
\fi

\begin{document}

\title{Iran's January 2026 Internet Shutdown: Public Data, Censorship Methods, and Circumvention Techniques}

\author{Giuseppe Aceto}
\orcid{0000-0002-4445-6259}
\affiliation{
  \institution{University of Napoli Federico II}
  \department{Department of Electrical Engineering and Information Technology}
  \city{Napoli}
  \country{Italy}
}
\email{giuseppe.aceto@unina.it}

\author{Valerio Persico}
\orcid{0000-0002-7477-1452}
\affiliation{
  \institution{University of Napoli Federico II}
  \department{Department of Electrical Engineering and Information Technology}
  \city{Napoli}
  \country{Italy}
}
\email{valerio.persico@unina.it}

\author{Antonio Pescapè}
\orcid{0000-0002-0221-7444}
\affiliation{
  \institution{University of Napoli Federico II}
  \department{Department of Electrical Engineering and Information Technology}
  \city{Napoli}
  \country{Italy}
}
\email{pescape@unina.it}

\renewcommand{\shortauthors}{Giuseppe Aceto, Valerio Persico, and Antonio Pescapè}



\begin{abstract}

This paper analyzes the Internet shutdown that occurred in Iran in January 2026 in the context of protests, focusing on its impact on the country's digital communication infrastructure and on information access and control dynamics. 
The scale, complexity, and nation-state nature of the event motivate a comprehensive investigation that goes beyond isolated reports, aiming to provide a unified and systematic understanding of what happened and how it was observed.

The study is guided by a set of research questions addressing: 
the characterization of the shutdown via the timeline of the disruption events and post-event ``new normal''; 
the detectability of the event, encompassing monitoring initiatives,  measurement techniques, and precursory signals; 
and the interplay between censorship and circumvention, assessing both the imposed restrictions and the effectiveness of tools designed to bypass them.

To answer these questions, we adopt a 
multi-\-source, multi-\-perspective
methodology that integrates heterogeneous public data, primarily from grey literature produced by network measurement and monitoring initiatives, complemented by additional private measurements. 
This approach enables a holistic view of the event and allows us to reconcile and compare partial observations from different sources.

\end{abstract}

\maketitle

\section{Introduction}
\label{sec:intro}
This paper investigates the Internet shutdown that occurred in Iran in January 2026. 
The event, which took place in the context of widespread protests, deeply modifyed the Iranian connectivity to the Internet, affecting the country's digital communication infrastructure, with significant impact on information access and censorship dynamics.

The scale and complexity of the communication systems involved in the event, and the nature of the actor (nation-state level) fuel the interest in the analysis of what happened, of the techniques used to enact it, how these have been detected and monitored, and their interaction with censorship circumention practices.

We aim to characterize the shutdown, understand how it was implemented and detected, and assess its implications in terms of control and circumvention.

To guide the analysis, we formulate six research questions, drilling down in this complex topic:

\begin{itemize}
    \item[RQ1] \emph{What is known}---considering official and reliable sources---about the events affecting Iran’s digital communication infrastructure between December 2025 and January 2026, in correspondence with widespread protests across the country?
    More specifically, which components of the digital infrastructure were disrupted, for how long, and with what effects?
    
    \item[RQ2] \emph{Which ``new normal'' was established} after the end of the shutdown? How control changed w.r.t. before the shutdown?

    \item[RQ3] \emph{Which precursory symptoms} manifested, if any? Was the shutdown foreseeable?
    
    \item[RQ4] \emph{Which monitoring initiatives} contributed technical analyses of the event? What are the nature, the aim and capabilities of these initiatives and how they contributed to the understanding of the event?

    \item[RQ5] \emph{How informative observations could be drawn?} Which measurement platforms and techniques were used? Which data are available? Which limitations techniques and data present?

    \item[RQ6] \emph{Which censorship-circumvention tools} had the potential to bypass the imposed restriction and were they effective during the Internet shutdown?
    How effective these tools have been?
    
\end{itemize}

These questions are answered in the course of this work, in the form \emph{ARQn} (Answer to Research Question \emph{n}). 
To answer these questions, we adopt a \emph{multi-source approach} integrating all publicly available network monitoring information.
To this aim we systematically collected and analyzed all available sources on the censorship event, also complementing it with additional private data from a transit provider.
Due to its recency, these sources consist primarily of grey literature, including reports and technical analyses produced by monitoring initiatives.

The article is organized as follows.
Section~\ref{sec:background} presents the background including the iranian network infrastructure, the known interference practices implemented, the initiatives monitoring them, the available circumention tools to bypass such censorship efforts, and the history of recent distruptions.
Section~\ref{sec:timeline} 
integrates the contributions of the monitoring initiatives considered and provides a unified timeline of events, from the earliest signs of network disruptions to the outage and subsequent restoration, until the new disruption observed at the end of February with the military operations impacting the country.
Section~\ref{sec:sources_and_initiatives} presents in detail the data sources and the monitoring initiatives that informed the analysis of the 2026 shutdown, examining their institutional nature, methodological approaches, and resulting outputs.
Section~\ref{sec:circumvention}
examines the mechanisms available to circumvent censorship in response to the shutdown.
Finally, Section~\ref{sec:conclusion} concludes the paper by providing a summary of the main findings and a concise answer to the research questions.

\section{Background}
\label{sec:background}
To contextualize the events under study, this background section is structured in five parts. 
First, we describe the organization and structural characteristics of the Internet in Iran, outlining the governance features that shape its connectivity.
Second, we introduce the technical aspects of the interference mechanisms that the iranian network are currently known to be subjected to.
Third, we summarize the main circumvention tools available to bypass govern censorship.
Four, we provide an overview of the main monitoring initiatives that have contributed to observing and documenting network conditions in the country. 
Finally, we review the major network disruptions in Iran, in order to situate the 2026 shutdown within its broader historical and operational context.

\subsection{Iran's national network infrastructure}

For several decades, Iran's connectivity to the global Internet has been mediated through two principal international gateways~\cite{madory_iranian_shutdown_2026}:
\begin{enumerate*}[label=\textit{(\roman*)}]
\item the Telecommunication Infrastructure Company (TIC) 
(AS49666, formerly AS12880 and AS48159); and
\item the Institute for Research in Fundamental Sciences 
(IPM) (AS6736).
\end{enumerate*}
IPM established the country's first international Internet connection in the 1990s and since then primarily serves academic and research institutions. 
Subsequently, TIC expanded its role to include Internet service provision and has since become the main carrier of Iran's inbound and outbound Internet traffic.
This notwithstanding, IPM has maintained a technically distinct connection to the global Internet. 

According to available reports, all major telecom operators in Iran are one way or another controlled by the government.%
\footnote{\urldate{https://ainita.net/ownership-in-irans-telecom-sector/} }
Iran's online environment is known to be one of the most restrictive worldwide, with the authorities systematically increasing the cost and technical barriers to accessing the global Internet in order to steer users toward a domestically controlled network, while employing pervasive censorship, surveillance, content manipulation, and extralegal repression~\cite{FreedomHouse2024IranFOTN}.
At the core of this strategy lies the \textit{National Information Network (NIN)}, a centralized, state-operated digital infrastructure officially framed as a vehicle for cybersecurity and digital sovereignty~\cite{afpc2025riseOfNIN}. 
Amid mounting internal and external pressures, the NIN has evolved into a central instrument not only for managing unrest but also for preemptively suppressing it~\cite{afpc2025riseOfNIN}.

\subsection{Network interference practices}
Studying the censorship practices enforced on the Iranian network is hindered by the limited availability of in-country measurement vantage points.
Moreover, reported regional and ISP-level differences make it difficult to draw broad conclusions.
Indeed, anecdotal accounts and scientific studies (the most recent and technically detailed being \citet{tai2025irblock} and \citet{lange2025iranconsistencies}, here used as main references) show that Iran relies on an extensive censorship system 
(the so-called \emph{Great Firewall of Iran}, GFI).

The adopted censorship mechanisms have evolved substantially since the first systematic investigation in 2013~\cite{aryan2013internet}, continually incorporating new and more advanced blocking techniques as emerging protocols and applications appear,
but still relying also on the analysis of unencrypted protocols. %
According to recent studies,
the GFI is built on layered filtering methods, including DNS, HTTP, HTTPS, and protocol-level blocking. %
More specifically, the GFI \emph{intercepts DNS queries}, and for censored domain names responds with private addresses (that are routed inside the NIN) corresponding to block pages;
\emph{inspects HTTP fields} for keywords and domains in the first data packet of an established TCP connection (stateful inspection), returning a \texttt{403 forbidden} HTTP code if censored;
\emph{inspects SNI field of HTTPS handshake} for domains in the first data packet of an established TCP connection (stateful inspection), injecting \texttt{RST} packets;
finally, UDP traffic (of specific significance because of increasing QUIC adoption, and also related to VPN usage) is monitored and selectively filtered.

The GFI exhibits bidirectional behavior
(similar to censoring systems in China and Turkmenistan)
i.e. it 
takes interference actions on both incoming and outgoing traffic, regardless of the traffic's origin. %
This allowed active probing of the censoring mechanism from outside the NIN.
Moreover, 
experimental analyses suggest
a nuanced and possibly hierarchical structure of censorship deployment:
while the GFI operates as a centralized system, certain ASes may be granted exemptions or are subject to less stringent censorship measures.
Some ASes
apply nearly complete censorship across DNS, HTTP, and UDP protocols, 
whereas others, 
show negligible or no censorship. %
These accounts offer several snapshots in time of a highly dynamic infrastructure of communications control, therefore they do not necessarily apply to the shutdown of January 8th 2026.
Moreover, several analyses are built on measurement techniques and postprocessing steps that are not readily available to the public.

\subsection{Circumvention of Internet Censorship}
\label{sec:bg_circumvention}
A recent study~\cite{vafa2025learning} showed Virtual Private Networks (VPNs) among the most widely used solutions for circumventing network censorship.
This category includes several tools (either repurposed or specifically designed to circumvent censorship), united by the common overall mechanism that creates an encrypted tunnel between a user's device and a remote server located outside the censored network, preventing the access network operators from observing the final destination of traffic or tampering with DNS queries. %

A notable architecture, initially designed for privacy enhancement, is represented by Tor (The Onion Router), which provides anonymity and censorship resistance through a network of volunteer-operated relays.
Tor routes traffic through multiple encrypted hops, making it difficult for both intermediaries and observers to determine either the source or the destination of communications, while presenting the user a TCP proxy (on Android devices, a VPN-like connection).
To counter increasingly sophisticated DPI systems, Tor has also introduced \emph{pluggable transports} designed to obfuscate Tor traffic so that it resembles ordinary protocols, thereby reducing the likelihood of detection by national filtering systems. %
More recently, to counteract the server enumeration attacks that mostly affects VPNs, introduced the \emph{Snowflakes} architecture%
\cite{bocovich2024snowflake}:
 a large number of ultra-light, temporary proxies (``snowflakes''), which accept traffic from censored clients using peer-to-peer WebRTC protocols and forward it to a centralized bridge.
 The temporary proxies can be implemented in JavaScript, in a web page or browser extension (much cheaper to run than a traditional proxy or VPN server).

Psiphon represents another widely deployed censorship circumvention platform designed specifically for users in heavily filtered networks, and its effectiveness in Iran has been analyzed during the Iranian elections in 2013 and 2016\cite{deibert2019censors}.
Unlike Tor’s decentralized volunteer infrastructure, Psiphon uses a managed network of proxy servers combined with multiple tunneling protocols, including VPN, SSH, and HTTP proxy technologies.
The system dynamically selects connection methods and employs traffic obfuscation to evade both DNS manipulation and DPI-based blocking.
Its centralized design emphasizes accessibility and rapid adaptation to evolving censorship techniques.
 
\subsection{Observing Internet Connectivity in Iran}

Besides field reports, consisting in observations, findings, or activities recorded by people directly experiencing the effects of the control mechanisms impacting the network,
the analyses performed at larger scales rely on different data sources, such as routing connectivity, active probing, and passive data-traffic observation.

Over the years, numerous monitoring initiatives have contributed to the analysis of the state of the Internet in Iran, including academic projects (IODA), private-sector platforms (Cloudflare Radar, Kentik, Whisper), independent observatories (NetBlocks' Internet Observatory, OONI), and initiatives focused on human-rights (FilterWatch).
More details on available data sources and monitoring initiatives are reported in Section~\ref{sec:data_sources} and Section~\ref{sec:initiatives}, respectively.

\subsection{History of Previous Network Disruptions in Iran}

Iran has repeatedly resorted to Internet shutdowns in response to episodes of domestic unrest, deploying measures that variably affect the NIN and the country's connectivity to the global Internet. 
Since 2019, three major nationwide disruptions have been documented, as discussed below.

\customSubsectionWithDate{Bloody November Shutdown}{November 2019, $\approx$ 6 days}
This was the first shutdown observed by IODA and was primarily implemented through the withdrawal of BGP routing announcements. Although Iranians experienced a near-total blackout, variations across IODA data sources and OONI measurements indicate that disconnection mechanisms differed by ISP and that limited access (particularly to the NIN) persisted during the disruption~\cite{meng_bischof_dainotti_iran_shutdowns_2026}.
It involved both cellular and fixed-line networks.

\customSubsectionWithDate{Women, Life, Freedom Movement Shutdown}{September 2022, $\approx$ 2 weeks}
Authorities imposed recurring nightly shutdowns targeting mobile networks
(the strategy visible in IODA routing and active probing data) while keeping fixed-line connections online to reduce the broader economic and political costs of a full blackout~\cite{ioda2022iran, meng_bischof_dainotti_iran_shutdowns_2026}. 
These curfews, totaling roughly 100 hours, were accompanied by additional measures such as application and protocol blocking to further restrict information flows and access to circumvention tools~\cite{ioda2022iran}.

\customSubsectionWithDate{Stealth blackout during the Twelve-Day War with Israel}{June 2025, $\approx$ 2 weeks}
Unlike previous disruptions, the June 2025 shutdown did not rely on withdrawing BGP routes, thereby preserving the outward appearance of normal connectivity for conventional monitoring systems. 
Instead, authorities implemented a centralized filtering regime at the national gateway, combining DNS poisoning, protocol \emph{allowlist}, and Deep Packet Inspection (DPI) to block foreign services and neutralize circumvention tools while maintaining selected domestic access. 
The government justified the restrictions as a means of deterring cyberattacks from Israel.
The shutdown involved both cellular and fixed-line networks.
This strategy reduced Iran’s international Internet traffic by approximately 90\% without fully disconnecting the country~\cite{filterwatch2025stealth, madory_iranian_shutdown_2026}.
Yet coordinated and diversified responses from the Internet freedom community enabled millions of Iranians to retain partial global connectivity.

\section{2026 Iran's Disruption: Lines of Evidence and Analyses}
\label{sec:timeline}

\begin{figure*}
    \centering
    \includegraphics[width=1\linewidth, trim={0 21cm 0 0}, clip]
    {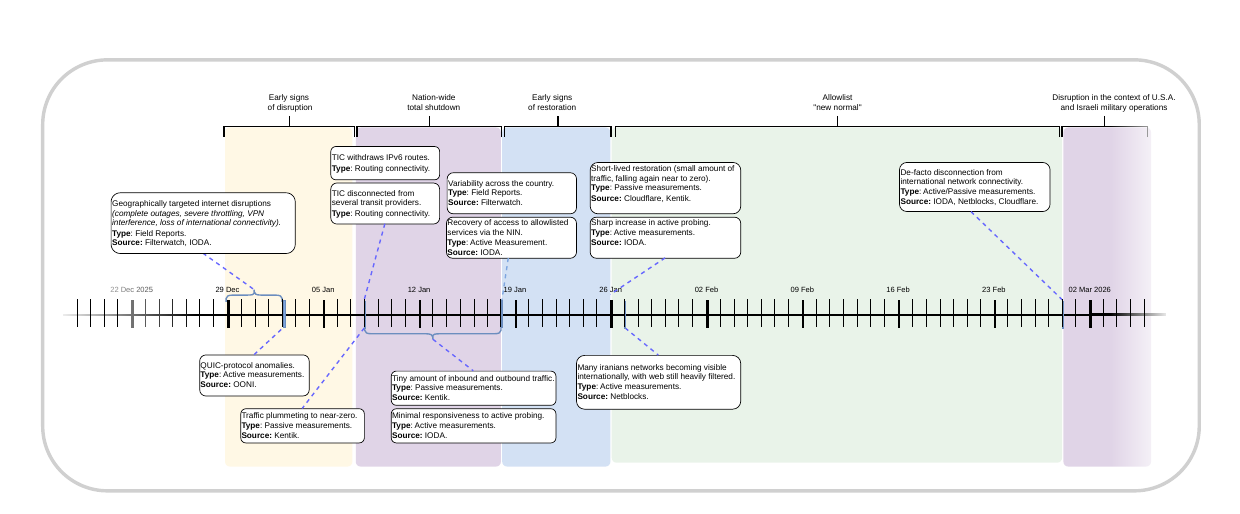}
    \caption{Timeline of main registered disruption events.}
    \label{fig:timeline}
\end{figure*}

\begin{figure*}
    \centering
    \includegraphics[width=0.9\linewidth]{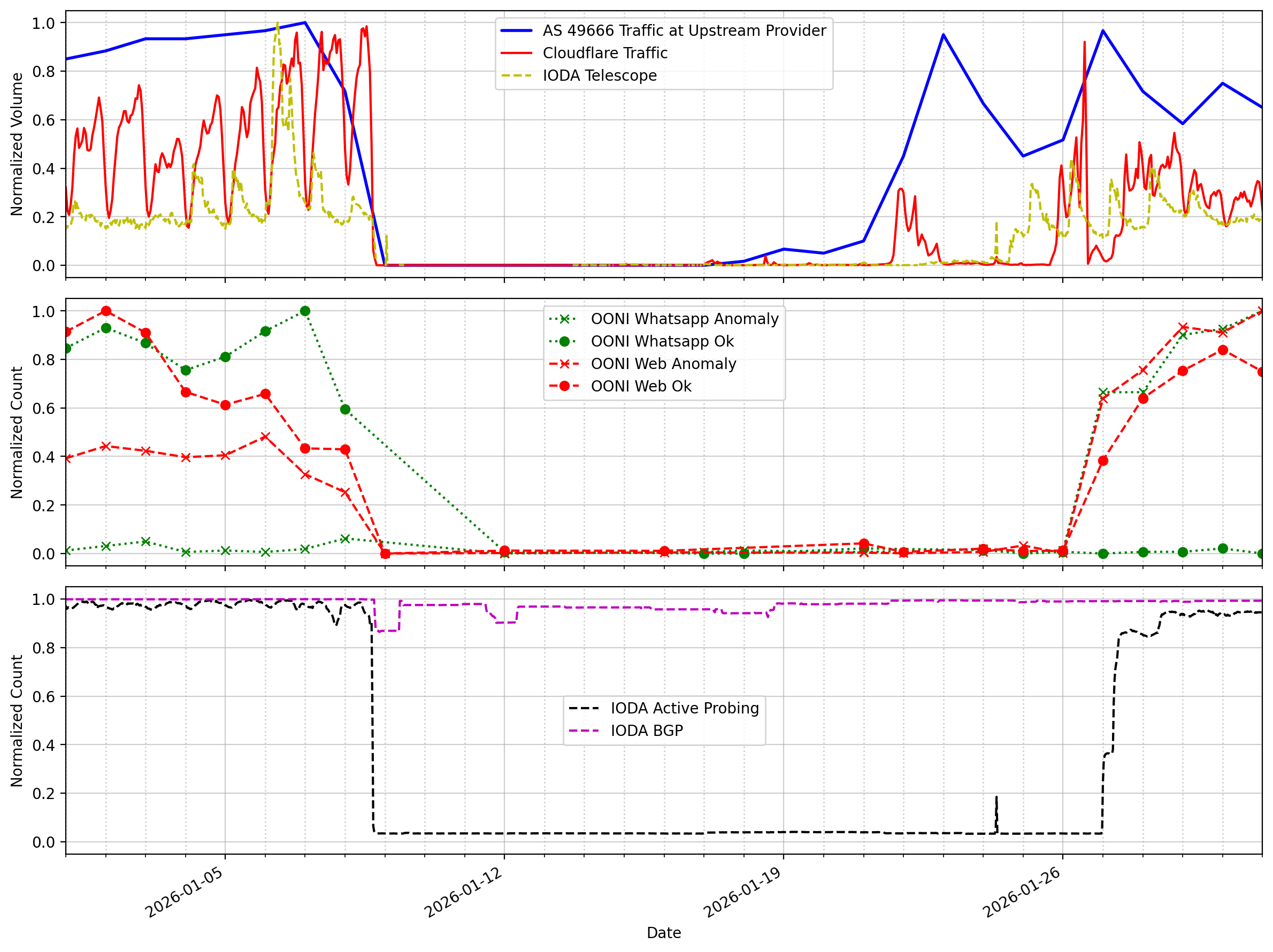}
    \caption{Synoptic monitoring data: 
    passive measurements of traffic and telescope data (top); OONI active probing (middle); IODA active probing and BGP data (bottom). }
    \label{fig:observations}
\end{figure*}

Figure~\ref{fig:timeline} presents a timeline that summarizes the observations of events recorded in relation to the disruption of Internet connectivity in Iran. 
This timeline is substantiated by monitoring data we collected from different public sources and complemented with the information provided by a transit provider under NDA.
Figure~\ref{fig:observations} constitutes an integrated overview derived from the set of available monitoring sources. 
It combines passive measurements (Cloudflare Radar data, traffic observed by a transit provider, and the IODA telescope), active measurements (IODA and OONI), and BGP routing information. 
The observation period covers the whole month of January 2026, encompassing the shutdown and surrounding days.
The different data sources present a broadly consistent picture of the event. 
Passive and active measurements largely agree on the timing and progression of the disruption. 
The small visible temporal discrepancies observed in the passive traffic analysis from the transit provider can be attributed to the coarser time granularity of the data, which is aggregated using daily averages. 
In contrast, BGP-level observations do not show corresponding anomalies, indicating that the disruption was not implemented through BGP routing changes but rather through mechanisms operating at other layers of the network.
In the following, the related details are provided.

\customSubsectionWithDate
{Early signs of disruption}
{Dec 29th, 2025--Jan 7th, 2026}
Starting on December 29, field reports gathered by Filterwatch~\cite{filterwatch2026december2025} and IODA~\cite{meng_bischof_dainotti_iran_shutdowns_2026}
indicated the onset of a geographically targeted Internet disruptions in Tehran and several provincial cities, with complete outages, severe throttling, and VPN interference concentrated in commercial, administrative, and protest-prone areas, while other nearby locations remained partially functional~\cite{filterwatch2026regional}.
Disruptions ranged from severe slowdowns to the loss of international connectivity, often restricting users to domestic services.
Mobile and fixed-line networks did not exhibit uniform behavior.
In several instances, fixed-line services were disrupted while mobile data remained active, and in others, access to (internal) NIN was preserved while international traffic was severed or throttled, with inconsistencies also observed among fixed-line providers regarding outages and performance degradation~\cite{filterwatch2026regional}.
The users experienced the Internet appearing connected, but with services and apps not actually loading.
Also VPNs resulted unstable.
According to these field reports~\cite{filterwatch2026december2025}, 
these phased and localized measures were implemented at granular network levels (e.g., ISP PoPs or cell towers) and enabled authorities to \emph{disrupt coordination without necessarily triggering the nationwide traffic drops} that would be visible in macro-level monitoring data.

On January 2, the OONI Observatory reported protocol-level anomalies affecting QUIC traffic on specific networks (the staggered appearance of these disruptions across operators suggests a phased implementation).%
\footnote{\urldate{https://x.com/OpenObservatory/status/2007508011812892743?s=20}}
Such initial evidences strongly suggested that while the 
network appeared as technically ``connected'' at the IP layer, disruption was shifted to higher layers.

\customSubsectionWithDate
{Nation-wide Total Shutdown}
{Jan 8th--Jan 18th 2026}
The first major development occurred on January 8 at 11:42 UTC when TIC (AS49666) started withdrawing its IPv6 BGP routes from its sessions, causing nearly all Iranian IPv6 routes to vanish from the global routing table within hours~\cite{madory_iranian_shutdown_2026}.
However, since IPv6 typically accounts for less than $1\%$ of total inbound traffic to Iran (source: Kentik's aggregated NetFlow data) the average user was unlikely to notice any impact~\cite{madory_iranian_shutdown_2026, whisper2026iranblackout}.

After a brief disruption, also IPv4 traffic began to sharply decline at 16:30 UTC (19:00 local time), falling until it had nearly ceased by 18:45 UTC. 
This was reflected in IODA Telescope data as a sudden drop, in Cloudflare metrics as a sharp nationwide traffic collapse~\cite{filterwatch2026regional,cloudflareRadarIran2026, meng_bischof_dainotti_iran_shutdowns_2026},
and traffic volume seen at an upstream provider of AS 49666
(see Fig.~\ref{fig:observations}, top).
A corresponding drop of active measurements can be found in OONI probing testing (see Fig.~\ref{fig:observations}, middle) 
and IODA active probing (see Fig.~\ref{fig:observations}, bottom).

In total, it took more than two hours to bring inbound and outbound connectivity to a near standstill~\cite{madory_iranian_shutdown_2026, meng_bischof_dainotti_iran_shutdowns_2026}.
At 19:00 UTC, TIC was observed disconnecting from several transit providers—including Russian state-owned Rostelecom (AS12389) and Gulf Bridge International (AS200612)—as well as from all of its settlement-free peering partners~\cite{madory_iranian_shutdown_2026, whisper2026iranblackout}, with the reduced amount of visible /24s ($\approx-12\%$) also highlighted by IODA~\cite{meng_bischof_dainotti_iran_shutdowns_2026}.
On the same day, Netblocks' data reported the total loss of connectivity on Iran Internet backbone (TCI provider)
affecting restive cities such as Kermanshah.
By 22:15 Iran time, Kentik data confirms the Internet reached a state of total blackout, with traffic plummeting to near-zero~\cite{filterwatch2026regional}.
Notably, the separation from the state-owned TIC has not shielded IPM from state-imposed censorship or monitoring measures~\cite{madory_iranian_shutdown_2026}.

Despite losing many BGP adjacencies for AS49666 (TIC), most Iranian IPv4 prefixes remained globally routed~\cite{madory_iranian_shutdown_2026} (see IODA BGP data in Fig.~\ref{fig:observations}, bottom).
The sharp decline in IPv4 traffic was therefore not due to reachability loss, but to filtering at the network edge.
IODA active-probing and BGP data illustrate this clearly (see Fig.~\ref{fig:observations}, bottom): 
active probing fell to zero as traffic was blocked, while routed IPv4 space in BGP remained largely intact 
($\approx$98.14\% after Jan 8th)~\cite{meng_bischof_dainotti_iran_shutdowns_2026}.
Hence, routing announcements were not the main mechanism used to generate the shutdown, highlighting the regime's more sophisticated approach to implementing shutdowns and controlling the information environment.

In the initial phase of the shutdown, authorities cut both international connectivity and domestic infrastructure, including the NIN, privileged SIM cards, and landline services.
By disabling external and internal communication channels alike, the state imposed an unprecedented level of isolation, even exceeding the wartime disruptions of 2025, and creating a near-total information vacuum~\cite{filterwatch2026_breakdown}.

After January 8, the Internet shutdown in Iran was not complete:
a tiny amount of traffic was still flowing in and out%
~\cite{madory_iranian_shutdown_2026}.
IODA measurements recorded minimal responsiveness to active probing ($\approx$3\%), likely reflecting either measurement artifacts or residual connectivity reserved for allowed users, such as government entities or services operating within Iran's state-controlled network perimeter \cite{meng_bischof_dainotti_iran_shutdowns_2026}.
Outside of very limited connectivity, digital human rights groups report severely limited access to the Internet both internationally and domestically.  
This includes limited, intermittent ability to make phone calls via landlines~\cite{meng_bischof_dainotti_iran_shutdowns_2026}.

\customSubsectionWithDate
{Early Signs of Restoration}
{Jan 18th--Jan 26th 2026}
On January 18, IODA, Cloudflare Radar, and Kentik reported early signs of Internet connectivity being restored in Iran \cite{meng_bischof_dainotti_iran_shutdowns_2026}. 
However, the apparent restoration of Internet connectivity observed proved to be short-lived, 
as the small volume of traffic quickly dropped back to near-zero (see Fig.~\ref{fig:observations}, top).
In accordance to this, IODA data showed synchronous reachability spikes for both Telescope and active probing on January 24th (see Fig.~\ref{fig:observations}, top and bottom).
A stronger recover was registered in the volumes of traffic at the upstream provider (see Fig.~\ref{fig:observations}, top).
After January 18
access was non-uniform across the country and varies by provider~\cite{filterwatch2026network}.

\customSubsectionWithDate
{Allowlist ``New Normal''}
{Jan 26th--Feb 28th 2026}
As of 10:00 PM (UTC) on January 26,
IODA observed a sharp increase in Active Probing witnessing the reachability of Iran's networks from the Internet (see Fig.~\ref{fig:observations}, bottom).%
\footnote{
\urldate{
https://bsky.app/profile/did:plc:3xessr3vu336mxean6zvfyjq/post/3mdfuqwxjpc2g
}
}
Such a change of connectivity status was also notified by Netblocks,
which on January 27th reported that many iranians networks were becoming visible internationally.%
\footnote{https://www.instagram.com/p/DUBHDR4DP8l/}

However, according to Netblocks, 
no \emph{return to normal} was registered: 
web was still heavily filtered on an \emph{allowlist} basis.
Despite this statement is not explicitly supported by shared evidences,
OONI data do provide experimental proof of this, especially regarding Whatsapp instant messaging.
Fig.~\ref{fig:observations} (middle) shows that before the shutdown, Whatsapp was generally accessible (low counts of anomalies), 
while a complete block is evident in the ``new normal'' phase.
Similarly, OONI Web tests reveal a significant increase of anomalies with respect to before the shutdown.

Traffic data from Kentik and IODA confirms this is a ``low-volume'' restoration. 
Traffic remains at only 25\% of pre-shutdown levels, further proving that the vast majority of the global Internet remains inaccessible to the general public~\cite{filterwatch2026network}.
Cloudflare data indicate that overall traffic volume, which had been steadily increasing since the end of January, experienced a marked decline around February 11~\cite{cloudflareRadarIran2026} (not evident from IODA dashboard). 
Following this abrupt drop, traffic levels stabilized at approximately 50\% of those observed in the immediately preceding period.

Focusing on the accessible services,
according to FilterWatch a specific set of international platforms and services have been allowlisted to provide a baseline of digital functionality, including 
search engines, communication tools, app stores, AI tools, gaming, and navigation.%
\footnote{
Service list: Google, Bing,
Google Meet, Gmail, Outlook,
Play Store, App Store, Apple, Samsung website,
ChatGPT, GitHub,
PlayStation, and
Google Maps~\cite{filterwatch2026network}.
}
Other social media and messaging platforms%
\footnote{
Instagram, Telegram, YouTube, and X.
}
remained accessible only via circumvention tools and continue to experience instability \cite{filterwatch2026network}. 
Others that were previously reachable%
\footnote{
WhatsApp and LinkedIn.
}
deteriorated to comparable restrictive conditions~\cite{filterwatch2026network}.
Also, user reports on January 27 indicate that a higher number of VPNs became functional compared to the initial days of the Internet shutdown, when most circumvention tools were rendered useless~\cite{filterwatch2026network}.

This situation reflects Iran's ``allowlist'' policy: 
a model akin to the Chinese Great Firewall that permits access only to approved users or services while blocking all others \cite{meng_bischof_dainotti_iran_shutdowns_2026}.
Field reports witness that since January 15, international connectivity in Iran has functioned as a tightly monitored privilege rather than a universal service. 
A pilot ``tiered Internet'' program at the Tehran Chamber of Commerce requires applicants to undergo in-person identity verification, register fixed IP addresses, and sign a written pledge not to ``misuse'' access~\cite{filterwatch2026network}.

\customSubsectionWithDate{Disruption in the context of U.S.A. and Israeli military operations}{from Feb 28th}
In the context of U.S.A. and Israeli military operations,
on February 28th IODA reported that Iran was cut off from the global Internet beginning at approximately 07:00 UTC, reflecting a de-facto disconnection from international network connectivity,
in line with the disconnection implemented during the Jun 2025 conflict.%
\footnote{\urldate{https://bsky.app/profile/did:plc:3xessr3vu336mxean6zvfyjq/post/3mfwfpezy6c2e}}
Similarly, Netblock's Internet-observatory data revealed that Iran was (and still is at the time of writing) experiencing an almost complete Internet shutdown, with national connectivity reduced to approximately 4\% of typical levels.%
\footnote{\urldate{https://mastodon.social/@netblocks/116147264437940657}}
A coherent assessment is provided by Whisper.
After 24 hours, the overall country connectivity was estimated at approximately 1\% of typical levels.%
\footnote{\urldate{https://mastodon.social/@netblocks/116149846834309209}}
Also According to the data from Cloudflare Radar, the traffic trends reveal near-total disconnection ($<0.1\%$ maximum level registered in the previous period).%
\footnote{\urldate{https://radar.cloudflare.com/ir?dateRange=28d}}

The analyses above allow us to answer some of the research questions, as follows.

\ansRQ{What is known about Iran's 2026 Internet shutdown}{1}
The events that occurred in the final days of 2025 and in January 2026 provide a new example of the Iranian State's interference in, and control over, the country's digital infrastructure. 
During this period, access to Internet-based services was rendered unavailable.
The shutdown lasted approximately ten days (8--18 January),
consistently with the other nationwide disruptions documented since 2019.
Although evidence of disconnections at the BGP level can be observed, these are limited both in scope (primarily affecting IPv6 routes) and in duration. 
Even during the shutdown period, BGP connectivity appears to be largely restored, revealing that the blocking mechanisms were implemented at higher layers of the protocol stack.

\ansRQ{Which ``new normal'' was established}{2}
After the shutdown, available measurements data reveal a drop in traffic volumes with respect to before, as well as increasing selective impairment of web and applications access.
Field reports explain that service restoration followed a selective approach, based on allowlist mechanisms and authorization procedures, potentially including in-person identification requirements. 
This approach
effectively embeds surveillance into access itself in a way that is hard to detect via network measurements from outside the national network. 
The described scenario lasted until February 28th,
when a total shuthdown was enacted again in correspondence of U.S.A. and Israeli military operations.

\begin{figure*}
    \centering
    \includegraphics[width=0.9\linewidth]{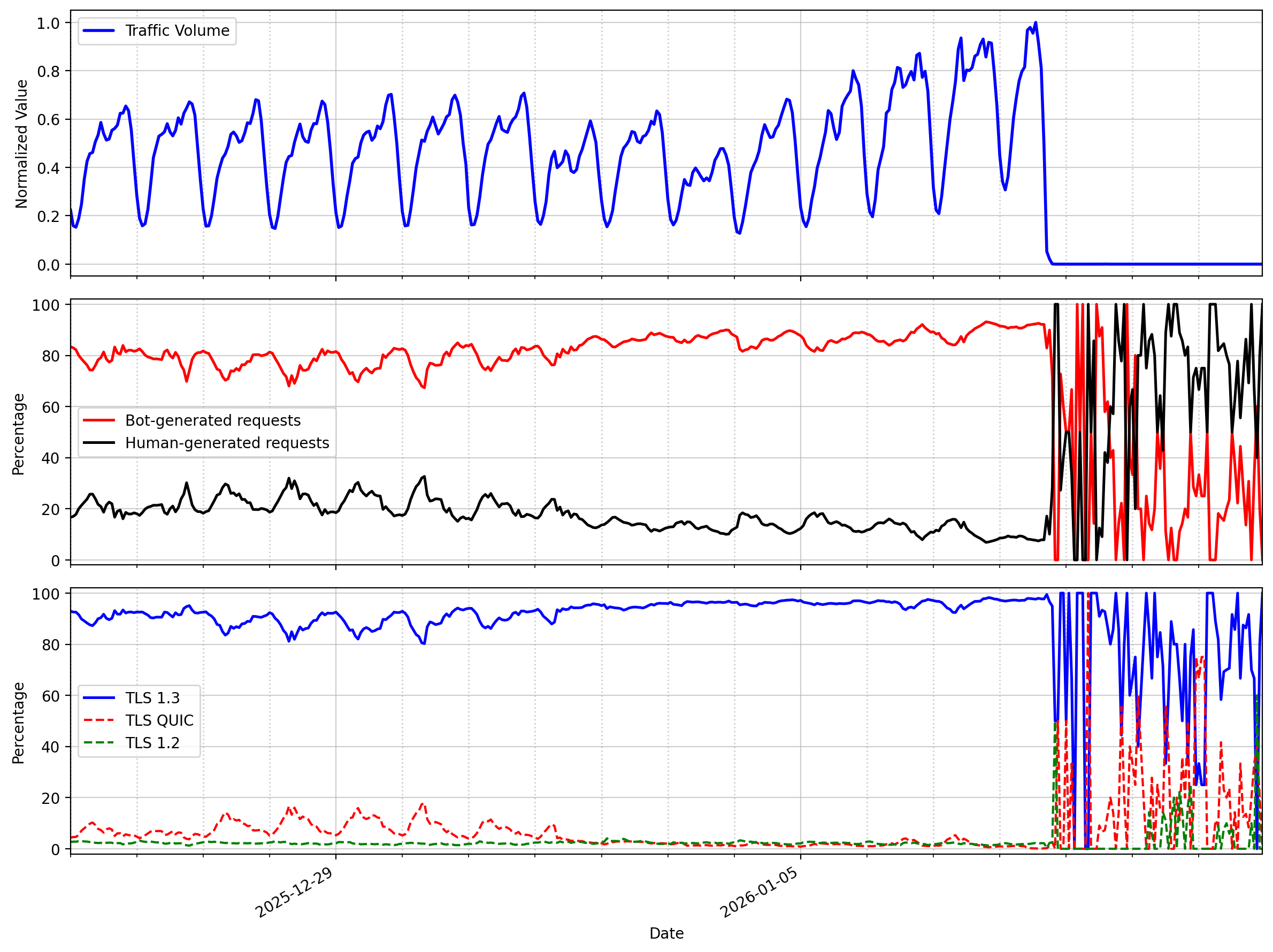}
    \caption{Pre-shutdown clues in bot- vs.~human- generated traffic and TLS protocol shares (Data Source: Cloudflare Radar).}
    \label{fig:preshutdown}
\end{figure*}

\ansRQ{Which precursory symptoms manifested}{3}
Before the shutdown, several field reports of connectivity disruption or impairing have been mentioned by human rights associations, on-alert because of the unrest and protests happening in several Iranian cities.
The localized and time-limited nature of this kind of interference makes it hard to detect with nation-wide aggregated monitoring techniques.

Some considerations can be proposed, regarding network patterns detectable starting on January 1st (noting that Jan 2nd has been a national holiday for Iran).
First, the web traffic originated from Iran towards the Cloudflare CDN and marked \emph{likely human-generated}%
\footnote{
Cloudflare tells human operators from automated web scraping by means of several proprietary heuristics, including to a large extent a supervised ML algorithm, see:
\urldate{https://developers.cloudflare.com/bots/concepts/bot-score/}.
}
(Fig.~\ref{fig:preshutdown}, top)
sees a significant reduction (bottoming on January 4th to about $35\%$ from the reference maximum).
By inspecting the bot-vs-human ratio
(Fig.~\ref{fig:preshutdown}, middle) emerges that bots increase their share from January 2nd on;
a similar phenomenon at the same time can be seen considering the share of TLS~QUIC
(Fig.~\ref{fig:preshutdown}, bottom)%
, which significantly shrinks with respect to other TLS versions, also losing the daily pattern typical of human activities.
Thanks to field reports, this human-selective traffic reduction can be related to unreliable connections (including possibly throttling), which frustrate human users (thus amplifying the connectivity impairment) while not (overly-) affecting automated traffic.

\section{Data Sources and Monitoring Initiatives}
\label{sec:sources_and_initiatives}
In the following, we present the data sources with the related monitoring approaches at the basis of the analysis of the connectivity in Iran (Section~\ref{sec:data_sources})
and detailed the monitoring initiatives that have provided effort in such analysis (Section~\ref{sec:initiatives}).
\subsection{Data sources}
\label{sec:data_sources}

The Iranian regime does not tend to disclose data regarding the management mechanisms of its national digital communication infrastructure. Consequently, information for outage analysis must be obtained through monitoring and telemetry approaches that mostly do not rely on the cooperation of the networks under observation.

Visibility on \textbf{global routing connectivity} represents a major source of information to analyze how the iranian networks are connected to the Internet.
RouteViews and RIPE RIS are the two main publicly accessible sources of information about BGP data.
They both take advantage of route collectors strategically deployed at IXPs and partner networks worldwide, where they peer with ISPs, academic networks, backbone providers, and hyperscalers to capture diverse, multi-perspective views of the global routing table.

While BGP data offers a high-level view of the Internet's fundamental connectivity structure, it is insufficient to uncover or explain more complex network management strategies.
To address this limitation, \textbf{active probing} (i.e., the injection of measurement traffic into the network to 
solicit some sort of responses and/or to observe how the traffic is handled) is used to complement and enrich the analysis.
Notably, obtaining reliable in-country vantage points for active measurements is challenging due to sanctions, strict regulations, and risks to local volunteers.
As a result, researchers often rely on external probing toward in-country services, although sending large volumes of probes to publicly accessible servers also raises concerns~\cite{tai2025irblock}. 

These insights can be further enriched through \textbf{passive data traffic observations} from privileged vantage points. 
Such vantage points may include upstream providers of Iranian networks (specifically, of the gateway providers TIC and IPM) where traffic volumes and related variations can be easily captured; 
or CDN infrastructures 
that have nodes within the country
(the latter also enabling inference on traffic characteristics
e.g., the analysis of request–response patterns
or human vs. automated activity). 
As a particular case of passive analysis,
it is worth to mention \textbf{darkspace traffic analysis}
that monitors traffic directed to routable but unassigned IP addresses (an evidence or byproduct of scanning and attacks).
A network telescope enables the detection of structural changes in a country's connectivity and network behavior, serving as a useful tool to identify nationwide blackouts, analyze disconnection events, observe systemic effects of censorship or government interventions.

\subsection{Monitoring Initiatives}
\label{sec:initiatives}
\textbf{IODA}%
\footnote{\urldate{https://ioda.inetintel.cc.gatech.edu/}}
monitors the connectivity of global Internet infrastructure providing data at country, subnational, and ISP/AS levels and allows to assess the severity of disruptions.
The project is run by the \emph{Internet Intelligence Lab}, an academic research lab at the School of Computer Science, College of Computing of Georgia Tech;
it was initially developed at the Center for Applied Internet Data Analysis (CAIDA), in the UC San Diego, and has been funded by the U.S.A. NSF and OTF organizations.
It relies on three main measurement types: Border Gateway Protocol (BGP) data (from RIPE NCC's Routing Information Service (RIS)
and the University of Oregon’s RouteViews project), active probing (4.2 million /24 address blocks are probed at least once every ten minutes), and network telescope observations (analysis of unsolicited background traffic).
IODA's long-term perspective on global Internet connectivity offers insight into the increasing sophistication of the Iranian authorities' approaches to information control and Internet shutdowns within the country
via both technical reports~\cite{meng_bischof_dainotti_iran_shutdowns_2026} and an online dashboard%
\footnote{\urldate{https://ioda.inetintel.cc.gatech.edu/country/IR?from=1764609812&until=1770830612&view=view1}}.

\textbf{Cloudflare Radar}%
\footnote{\urldate{https://radar.cloudflare.com/}}
is a 
platform that presents insights into global Internet traffic, cybersecurity threats, and technology adoption trends. It is primarily powered by data from Cloudflare’s global  Content Distribution Network and aggregated, anonymized data from their 1.1.1.1 public DNS resolver.
The platform also incorporates external datasets in specific areas, including interconnection metadata from PeeringDB, Autonomous System population estimates from APNIC, geographic reference data from GeoNames, and Internet routing information from RIPE NCC RIS and RouteViews.
The platforms allows to focus the analysis on selected nations, including Iran~\cite{cloudflareRadarIran2026}.

\textbf{Kentik}%
\footnote{\urldate{https://www.kentik.com/}}
is a U.S.A. technology company that develops a cloud-based network intelligence and observability platform,
which ingests and correlates a wide range of telemetry sources to give visibility into networks. 
These include:
\textit{core network telemetry} 
(flow data exported from routers, switches, hosts, and cloud environments;
routing and path data such as BGP routing information;
device and interface metrics collected via SNMP and streaming telemetry;
synthetic test results generated by agents for active performance monitoring) 
as well as \textit{enrichment and contextual telemetry}
(GeoIP and threat intelligence feeds;
cloud provider flow logs;
business/operational metadata like DNS data, orchestration context,
user identity information from NAC/RADIUS/IPAM).
After the blackout, the data from Kentik were used to explain how the event developed and compare it with past internet shutdowns in Iran~\cite{madory_iranian_shutdown_2026}.

\textbf{The Internet Observatory} project%
\footnote{\urldate{https://netblocks.org/projects/observatory}}
uses measurement, classification, and attribution methods to identify internet disruptions, censorship, and cyberattacks affecting critical infrastructure.
It supports the reporting activities of 
\textit{NetBlocks}%
\footnote{\urldate{https://netblocks.org/}}%
, an independent, non-partisan organization that monitors global Internet connectivity, with a focus on digital rights, cybersecurity, and Internet governance.
Its monitoring approach emphasizes impartiality, accuracy, and methodologies designed to operate at global Internet scale.
The program combines first-party and third-party data sources to provide an integrated view of connectivity and platform availability during events such as crises, elections, civic actions, and natural disasters that may affect human rights and democratic processes.
Sources are 
classified by telemetry type, such as connectivity measurements (e.g., ICMP), traffic data (e.g., web, DNS, or flow aggregates), probe-based measurements (e.g., active or user-driven tests), and routing metrics.
Since the start of the shutdown, NetBlocks has provided via social media at least daily updates on the status of the connectivity in Iran.

\textbf{FilterWatch}%
\footnote{\urldate{https://filter.watch/english/about-us/}}
is an Iran-focused digital rights initiative that operates as a project of the Miaan Group%
\footnote{\urldate{https://miaan.org/}}%
, a non-profit organization based in the U.S.A., focused on human rights in Iran and the wider Middle East.
Filterwatch provides legal, technical, research, and advocacy support, and analyzes and documents the human rights impacts of Internet censorship, surveillance, and infrastructure policies, particularly in relation to Iran’s NIN. 
The project provides data, analysis, and tools to support accountability, advocacy, and the protection of online privacy and freedom of expression~\cite{filterwatch2026network}.
Preminent data sources are of the \emph{field report} kind, with gathering of local information.
This kind of contribution complements (and is used jointly in the analyses) the network measurements from outside the national network.
Indeed ``tactical'' local disruptions are not visible in the aggregated traffic, and due to Carrier-Grade NAT architecture of the Iranian national network, measurement tools like ones by IODA or Cloudflare cannot easily detect sub-national scale throttling or jamming.
Following the internet blackout in Iran, FilterWatch released a series of technical reports \cite{
filterwatch2026_breakdown, 
filterwatch2026regional, 
filterwatch2026december2025, 
filterwatch2026network}, mainly supported by the network measurement from IODA, Kentik, and Cloudflare.

\textbf{OONI}%
\footnote{\urldate{https://ooni.org}}
(the Open Observatory of Network Interference)
is a project aimed at decentralized efforts in documenting internet censorship around the world since 2012.
Born as a sub-project of the U.S.A. non-profit organization \emph{Tor Project}%
\footnote{\urldate{https://www.torproject.org}},
since 2024 it is backed by a dedicated Italian non-profit association.
The project provides
Free/Open-Source software (\emph{OONI Probe} mobile and desktop app) to measure the blocking of websites, instant messaging apps, and censorship circumvention tools;
real-time analysis and open publication of censorship measurements from around the world, published as open data on a dashboard%
\footnote{\urldate{https://explorer.ooni.org/}};
research reports documenting censorship events around the world, in collaboration with more than $50$ research and human-rights advocacy partners.
By providing the active probing applications for smartphones and desktop OSes, the OONI project allows the continuous worldwide crowdsourcing of reachability measurements for selected services, offering a view from inside the potentially censored networks, thus offering (real-time and publicly) a viewpoint different from most censorship detection and reporting initiatives.
This data is at the basis of analysis reports which are frequently published on the project website and some scientific peer-reviewed papers.
Regarding Iran, OONI published (in collaboration with ARTICLE19, ASL19, and Small Media) a report%
\footnote{\urldate{https://ooni.org/post/iran-internet-censorship/}}
about blocking evidence \emph{between 2014 and 2017};
a multi-stakeholder report%
\footnote{\urldate{https://ooni.org/post/2022-iran-technical-multistakeholder-report/}
--- facilitated by the European Commission and U.S.A., coordinated by OONI and ISOC, and including besides OONI also IODA, Measurement Lab (M-Lab), Cloudflare, Kentik, Censored Planet, ISOC, Article19.}
about the censorship following the death of Jhina (Mahsa) Amini, \emph{between 16th September 2022 to 16th October 2022};
an analysis%
\footnote{\urldate{https://ooni.org/post/2025-iran-censorship-womens-rights/}}
in collaboration with \emph{Miaan Group}
on blocking of selected women's rights websites (with data analysis \emph{covering Feb. 1st, 2024 to Oct 1st, 2025}).
At the time of writing, no analysis has yet been published about Jan. 8th 2026 shutdown.

\newcommand{\cmark}{\ding{51}}
\newcommand{\pmark}{$\lozenge$}
\newcommand{\xmark}{$\cdot$}
\newcommand{\fpmark}{1st}
\newcommand{\tpmark}{3rd}

\begin{table*}[ht]
\centering
\caption{Monitoring initiatives that focused on 2026 Iran Internet shutdown.
}
\label{tab:monitoring-iran-2026-v1}
\small
\setlength{\tabcolsep}{4pt}

\begin{tabularx}{0.82\textwidth}{l c c cc | c c c c | ccc}
\toprule
\multirow{2}{*}{\textbf{Initiative}} &
\multirow{2}{*}{\textbf{Reg.}}&
\multirow{2}{*}{\textbf{Org.}}&
\multicolumn{2}{c}{\textbf{Data origin}} &
\multicolumn{4}{c}{\textbf{Data types / measurement methods}} &
\multicolumn{3}{c}{\textbf{Output}} \\
\cmidrule(lr){4-5}
\cmidrule(lr){6-9}
\cmidrule(lr){10-12}
& & & \textbf{1st} & \textbf{3rd}
& \textbf{BGP} & \textbf{Traffic } & \textbf{Probing} & \textbf{Telescope}
& \textbf{Tech.\ rep.} & \textbf{Social} & \textbf{Dashboard} \\
\midrule

Cloudflare Radar &
USA &
\iconInd &
\cmark & \cmark &
\tpmark & \fpmark & \xmark & \xmark &
\cmark & \xmark & \cmark \\

FilterWatch &
USA &
\iconONG &
\xmark & \cmark &
\tpmark & \xmark & \tpmark & \tpmark &
\cmark & \xmark & \xmark \\

IODA &
USA &
\iconAca &
\cmark & \cmark &
\tpmark & \xmark & \fpmark & \fpmark &
\cmark & \cmark & \cmark \\

Kentik &
USA &
\iconInd &
\cmark & \cmark &
\tpmark? & ? & ? & \xmark &
\cmark & \xmark & \xmark \\

NetBlocks &
UK &
\iconInd &
\cmark & \cmark &
\tpmark? & \fpmark & \fpmark & \xmark &
\xmark & \cmark & \xmark \\

OONI &
ITA\textdagger&
\iconONG &
\cmark & \xmark &
\xmark & \xmark & \fpmark & \xmark &
\xmark & \xmark & \cmark \\

Whisper &
USA &
\iconInd &
\xmark & \cmark &
\tpmark & \xmark & \tpmark & \tpmark &
\xmark & \xmark & \cmark \\

\bottomrule
\end{tabularx}

\vspace{0.6em}
\footnotesize
\textbf{Legend.}
\raggedright
\textbf{Initiative} = Monitoring initatives in alphabetical order;
\textbf{Reg.} = Registered office country;
\textbf{Org.} = Organization type: 
\iconAca~Academic, \iconONG~Non-Profit, \iconInd~Business;
\textbf{Data origin}: \textbf{1st}/\textbf{3rd} -party data collection;
\textbf{BGP} = routing and prefix visibility;
\textbf{Traffic} = aggregated traffic measurements;
\textbf{Probing} = active measurements;
\textbf{Telescope} = unsolicited background traffic;
\textbf{Tech. rep.} = technical report/analysis;
\textbf{Social} = social media updates;
\textbf{Dashboard} = public dashboard. \\
\textdagger~Initially founded in USA and funded mainly by USA private and public entities. \\
\end{table*}

\textbf{Whisper}%
\footnote{\urldate{https://www.whisper.security/}}
is a cybersecurity company founded in January 2025 that focuses on internet infrastructure intelligence. 
Following the Internet blackout in Iran on 8 January, Whisper made an online dashboard available~\cite{whisper2026iranblackout} to provide ongoing visibility into the country’s connectivity status. 
The platform aggregates data from multiple measurement and routing intelligence sources (IODA, RIPE, and Cloudflare) and presents near-real-time indicators describing the evolution of the disruption, together with analytical commentary on the technical mechanisms believed to have been employed. 
Among the exposed metrics are BGP path counts, hourly BGP activity, BGP UPDATE volume, OONI network measurement results, and prefix visibility by protocol (IPv4 and IPv6 prefix counts derived from RIPE RIS routing data). 
The dashboard also highlights structural aspects of Iran’s connectivity, including foreign transit providers connected to Iranian networks, internal ASN interconnection patterns and topology, and changes observed across affected networks. 
Particular attention is given to the behavior of IPM’s 26 downstream networks after the blackout, with Iranian address space grouped by ASN to illustrate differential routing responses.
Beyond descriptive monitoring, the analysis advances hypotheses framed in terms of a ``digital kill chain'' interpreting the event as a staged sequence of control actions targeting routing visibility, external reachability, and internal network segmentation.

\textbf{Additional data sources.}
We complemented the aforementioned data sources with a view of the aggregated and normalized traffic volume involving the Iranian network (AS49666) as seen at one international upstrem provider.
Due to NDA, we are not allowed to share further details on such traffic, but we report these data
due to their strong coherence with the rest of the available observations and since it constitutes a valid example of data sources that can shine more light on the analyzed event.

Table~\ref{tab:monitoring-iran-2026-v1}
summarizes the main monitoring initiatives that contributed to the analysis of the 2026 network shutdown in Iran, highlighting their country of registration, data and measurement methodologies, and produced outputs.

\ansRQ{Which monitorig initiatives contributed to the understanding of the events}{4}
What is currently known about these events affecting the network derives from a limited number of monitoring initiatives, whose outputs are publicly accessible in the form of online dashboards or technical reports. 
Most of these initiatives are U.S.A.-centric: 
they are typically legally based in the United States and, in the majority of cases, receive funding either from U.S.A. government sources or from private U.S.A. entities.
Despite the very high and well-established levels of expertise within academia and the private sector, as well as the significant efforts of non-governmental organizations—many of which are directly engaged in field operations—this structural configuration results in a concentration of both operational capacity and financial support within virtually a couple countries alone. 
An evident missing actor is the European Union, whose most-related monitoring project is \emph{Mapping Media Freedom}%
\footnote{\urldate{https://www.mappingmediafreedom.org/}}
which is not technically-focused, and is limited only to European countries.
Consequently, the continuity, scope, and methodological orientation of Internet monitoring activities remain significantly dependent on the political priorities, regulatory frameworks, and funding decisions of almost a single government.
While this does not diminish the professionalism or integrity of the actors involved, it does highlight an underlying systemic dependency that may affect long-term sustainability, pluralism of perspectives, and the global representativeness of the resulting datasets.

\ansRQ{How informative observations could be drawn}{5}
Most of the considered reports leveraged multiple viewpoints (combining multiple data sources, potentially originated at multiple vantage points), although not all the monitoring initiatives relied on their own managed monitoring infrastructures. 
Some primarily aggregate, curate, or reinterpret data produced by other entities. 
First-hand data are in fact provided by a relatively small set of primary sources, and raw datasets are not always publicly available.
These data sources can be broadly grouped into four categories: 
BGP data, passive traffic analysis, active probing measurements, and network telescopes (i.e., darknet monitoring infrastructures).
Importantly, large-scale monitoring platforms---by design---exhibit structural blind spots when confronted with layered, phased, or highly granular disruption events. 
In such cases, purely measurement-based approaches may fail to capture the full scope or timing of restrictions. 
The integration of field reports has therefore proven essential, particularly for the early detection of impending shutdown events.

\section{Circumvention response to the shutdown} 
\label{sec:circumvention}
Circumvention tools play a key role in reacting to Internet censorship activities.  
Indeed, VPNs and VPN-like tools as Tor and Psiphon have been widely used to circumvent Internet censorship~\cite{vafa2025learning},
and have been selectively subjected to censorship in their own turn.
During the analyzed shutdown, none of the circumvention techniques based on common Internet connectivity could work,
therefore other communication means have been explored, namely (illegal) satellite connections and peer-to-peer decentralized WLAN / PAN communications: these are described in the following sections.

\subsection{Satellite connection}
\label{sec:satellite}
The filtering of landline connection between Iran and the rest of the Internet has been circumvented through satellite access (specifically, with the \emph{Starlink} constellation).
On January 13th the USA financial news outlet Bloomberg published%
\footnote{\urldate{https://www.bloomberg.com/news/articles/2026-01-13/musk-offers-free-starlink-in-iran-as-internet-blackout-persists}}
news that Starlink access was provided for free to Iran citizens.
Indeed, despite possession of satellite communication technology is illegal in Iran, civil rights organizations have helped smuggling in the order of $50,000$ terminals in the country since 2022 protests%
\footnote{\urldate{https://www.nytimes.com/2026/01/15/technology/iran-online-starlink.html}}.
This in turn prompted the Iranian government to interfere with satellite link through jammming
(reported heavily happening from January 11th)%
\footnote{\urldate{https://filter.watch/english/2026/01/13/network-monitoring-january-2025-internet-repression-in-times-of-protest/}}
with up to $80\%$ packet loss%
\footnote{\urldate{https://filter.watch/english/2026/01/13/network-monitoring-january-2025-internet-repression-in-times-of-protest/}}
in some areas, for some of the time, while elsewhere the Starlink connections were less severely affected or not at all.

The cost of a satellite terminal, the lack of legal options to purchase them, and the risk associated with its possession (increased by the visibility of satellite ``dish'' antennas) all limit this option.
The recently deployed Direct-to-Cell~\cite{garcia2025direct}
\footnote{\urldate{https://www.spacex.com/updates/\#dtc-gen2-spectrum}}
Starlink technology would help in this regards, as common mobile phones could be used as terminals for satellite communications.
There are no reports of actual availability and usage of this technology during the analyzed event.
On the other hand, dependence on the goodwill of a private subject, currently running these services in a monopolistic condition, has already shown the reliability and (political) control risks in the case of Ukraine just a few months before%
\footnote{\urldate{https://www.reuters.com/investigations/musk-ordered-shutdown-starlink-satellite-service-ukraine-retook-territory-russia-2025-07-25/}}.

\subsection{Peer-to-peer messaging apps}
\label{sec:p2pmess}
If quasi-real time (``instant'') interactive communication is not a requirement, then long-distance digital communication is viable also in absence of Internet availability, by exploiting Personal-Area-Network technologies such as Bluetooth and delay-tolerant communication protocols.
This is the case of a set of apps reported to be used in Iran during the shutdown%
\footnote{\urldate{https://bitcoinmagazine.com/news/iranian-protestors-turn-to-censorship-resistant-freedom-tech-during-internet-blackout}}
namely
$Bitchat$%
\footnote{\urldate{https://github.com/permissionlesstech/bitchat/blob/main/WHITEPAPER.md}}
and its closed-source Iranian-localized fork \emph{Noghteha}, which has been downloaded more than $72,000$ times between Jan 8th and Jan 10 2026 from the mobile app market AppBrain%
\footnote{\urldate{https://www.appbrain.com/app/noghteha-mesh-messenger/com.filtershekanha.noghteha}[January 28th, 2026]
}.

This peak of downloads of \emph{Noghteha} testifies the sore need for secure and off-internet digital communication tools, but this need is hardly properly addressed by \emph{Noghteha} and \emph{Bitchat}.
Indeed, \emph{Bitchat} has been initially advertised as a secure communication app, but multiple reports in July%
\footnote{\urldate{https://www.supernetworks.org/pages/blog/agentic-insecurity-vibes-on-bitchat}}%
$^,$%
\footnote{Published on Jul 25, 2025 --- \urldate{https://saadkhalidhere.medium.com/why-bitchat-is-a-bad-idea-my-audit-found-critical-zero-days-1b126a45a2c5}}%
$^,$%
\footnote{Opened on bitchat github Issues list on Aug 1st, 2025 --- \urldate{https://github.com/permissionlesstech/bitchat/issues/376}}
revealed several critical security vulnerabilities, including two zero-days, which make the app ``dangerously insecure''.
Further concerns regard the closed-source forked version \emph{Noghteha} (the most used in Iran) whose security characteristics and development are completely opaque.
Both apps disclose that they didn't undergo an independent security audit and therefore should not be used in highly sensitive use cases.
Notably, \emph{Bitchat} has been also widely downloaded%
\footnote{\urldate{https://invezz.com/news/2026/01/14/bitchat-tops-uganda-app-stores-as-election-internet-blackout-drives-use/}}
in Uganda before the elections due Jan 15th 2026, as the government cut off the Internet%
\footnote{\urldate{https://reclaimthenet.org/uganda-imposes-nationwide-internet-shutdown-ahead-of-2026-election}}.
In this case, an Uganda IT taskforce allegedly%
\footnote{\urldate{https://nilepost.co.ug/news/315117/the-it-taskforce-behind-the-bitchat-block-ugandas-homegrown-cyber-defense-triumphs}}
succeded in blocking the Bluetooth communications and thus block the \emph{Bitchat} app communications, but no technical details nor independent validation of the claim has been provided.

\ansRQ{How effective censorship-circumvention tools have been}{6}
In the case of Iran, both Tor and Psiphon were blocked already before the shutdown, and no censorship detection has been possible during the shutdown itself (as very tight allowlist has been applied).
Notably, Tor resulted unfiltered from OONI tests after the shutdown, while Psiphon remained blocked afterwards.
The allowlist approach is effective against virtually all circumvention techniques that use the conventional local Internet connectivity:
this prompted for alternative means of communication, namely satellite connection and peer-to-peer WLAN and PAN messaging apps.
Both these approaches are virtually invisible to the monitoring tools used in this and other published analyses%
\footnote{
The decentralized nature of bitchat, using the \texttt{nostr} protocol, potentially allows for privacy-preserving activity monitoring with blockchain-based tracking dashboards such as \urldate{https://bitchatexplorer.com/}
}%
, so other approaches must be explored.

\section{Discussion and Conclusion}
\label{sec:conclusion}
The events of January 2026 constitute a further illustration of the Iranian authorities’ capacity to exert centralized control over the national digital infrastructure. 
The ten-day blackout (8–18 January) followed patterns observed in previous nationwide disruptions since 2019 and appears to have been implemented primarily through mechanisms operating above the BGP layer, rather than through sustained routing withdrawals.
The restoration of connectivity occurred through a selective and tightly controlled process based on allowlisting and authorization procedures. 
This model individualizes and conditions access to the network, embedding traceability and surveillance into connectivity itself and thereby significantly weakening guarantees of digital privacy.
Evidence suggests that network interference may have begun even before the official shutdown, through localized disruptions and degradations affecting human-generated traffic. 
These patterns (consistent with throttling or unstable connectivity) highlight the difficulty of detecting granular interference events using only nation-scale aggregated measurements.

The reconstruction of these events relies on a limited ecosystem of monitoring initiatives, many of which are institutionally and financially concentrated in a small number of countries, particularly the U.S.A. 
While these initiatives demonstrate high technical expertise and operational commitment, this concentration introduces structural dependencies that may influence agenda setting, long-term sustainability, and geographic coverage of monitoring activities.
Most available analyses combine heterogeneous data sources and observation vantage points, yet the underlying primary datasets originate from a relatively small set of infrastructures. 
BGP feeds, passive traffic measurements, active probing, and darknet monitoring collectively provide valuable insights, but measurement-based approaches alone exhibit blind spots when disruptions are layered, phased, or geographically localized, making \emph{field reports} an essential complementary source.

Finally, the shutdown demonstrated the effectiveness of \emph{allowlist} approaches against conventional circumvention tools operating over local Internet connectivity. 
As a consequence, alternative communication methods (such as satellite connectivity and peer-to-peer messaging applications based on local wireless networks) became particularly relevant, although these approaches remain effectively invisible to current large-scale monitoring frameworks and therefore require new methodological approaches for systematic observation.

\begin{acks}
We thank all the organizations and initiatives that collected, organized and made publicly available the monitoring data necessary for this analysis.
\end{acks}

\bibliographystyle{ACM-Reference-Format}
\bibliography{iran_blackout}

@article{garcia2025direct,
  title={Direct-to-Cell: A First Look into Starlink's Direct Satellite-to-Device Radio Access Network through Crowdsourced Measurements},
  author={Garcia-Cabeza, Jorge and Albert-Smet, Javier and Frias, Zoraida and Mendo, Luis and Azcoitia, Santiago Andr{\'e}s and Yraola, Eduardo},
  journal={IEEE Communications Magazine},
  year={2025},
  publisher={IEEE}
}

@article{deibert2019censors,
  title={Censors get smart: Evidence from Psiphon in Iran},
  author={Deibert, Ronald and Oliver, Joshua and Senft, Adam},
  journal={Review of Policy Research},
  volume={36},
  number={3},
  pages={341--356},
  year={2019},
  publisher={Wiley Online Library}
}

@inproceedings{vafa2025learning,
  title={Learning from Censored Experiences: Social Media Discussions around Censorship Circumvention Technologies},
  author={Vafa, Elham Pourabbas and Singhal, Mohit and Thota, Poojitha and Roy, Sayak Saha},
  booktitle={2025 IEEE Symposium on Security and Privacy (SP)},
  pages={1325--1343},
  year={2025},
  organization={IEEE}
}

@inproceedings{bocovich2024snowflake,
  title={Snowflake, a censorship circumvention system using temporary {WebRTC} proxies},
  author={Bocovich, Cecylia and Breault, Arlo and Fifield, David and Wang, Xiaokang and others},
  booktitle={33rd USENIX Security Symposium (USENIX Security 24)},
  pages={2635--2652},
  year={2024}
}

@inproceedings{aryan2013internet,
  title={Internet censorship in Iran: A first look},
  author={Aryan, Simurgh and Aryan, Homa and Halderman, J Alex},
  booktitle={3rd USENIX Workshop on Free and Open Communications on the Internet (FOCI 13)},
  year={2013}
}

@online{whisper2026iranblackout,
  title = {{The Blackout}},
  year = {2026},
  url = {https://state-of-iranblackout.whisper.security},
  note = {Interactive real-time report and analysis on Iran internet connectivity blackout in January 2026 (accessed: 2026-03-26)},
  organization = {Whisper},
  urldate = {2026-01-23}
}

@online{cloudflareRadarIran2026,
  title = {{Cloudflare Radar — Overview for Iran}},
  author = {{Cloudflare, Inc.}},
  year = {2026},
  url = {https://radar.cloudflare.com/ir},
  notexxx = {Internet traffic and connectivity metrics for Iran from Cloudflare Radar},
  organization = {Cloudflare},
  urldate = {2026-01-23}
}

@online{madory_iranian_shutdown_2026,
  author       = {Madory, Doug},
  title        = {From Stealth Blackout to Whitelisting: Inside the Iranian Shutdown},
  year         = {2026},
  month        = {Jan},
  day          = {21},
  note         = {\href{https://web.archive.org/web/20260122050624/https://www.kentik.com/blog/from-stealth-blackout-to-whitelisting-inside-the-iranian-shutdown/}{https://www.kentik.com/blog/from-stealth-blackout-to-whitelisting-inside-the-iranian-shutdown/} (Archived on 2026-01-22)},
}

@online{meng_bischof_dainotti_iran_shutdowns_2026,
  author       = {Meng, Amanda and Bischof, Zachary and Dainotti, Alberto},
  title        = {A Comparative Look at Internet Shutdowns in Iran: 2019, 2022, 2025, and 2026},
  year         = {2026},
  day          = {21},
  note         = {\href{https://web.archive.org/web/20260123175343/https://ioda.inetintel.cc.gatech.edu/reports/a-comparative-look-at-internet-shutdowns-in-iran-2019-2022-2026-and-2026/}{https://ioda.inetintel.cc.gatech.edu/reports/a-comparative-look-at-internet-shutdowns-in-iran-2019-2022-2026-and-2026/} (Archived on 2026-01-23)},
}

@inproceedings{tai2025irblock,
  title={{IRBlock}: A Large-Scale Measurement Study of the Great Firewall of Iran},
  author={Tai, Jonas and Sengottuvelavan, Karthik Nishanth and Whiting, Peter and Hoang, Nguyen Phong},
  booktitle={34th USENIX Security Symposium (USENIX Security 25)},
  pages={705--722},
  year={2025}
}

@article{lange2025iranconsistencies,
  title={I (ra) nconsistencies: Novel Insights into Iran’s Censorship},
  author={Lange, Felix and Niere, Niklas and von Niessen, Jonathan and Suermann, Dennis and Heitmann, Nico and Somorovsky, Juraj},
  journal={Free and Open Communications on the Internet},
  year={2025},
  issue = {1},
  pages = {7--12},
}

@online{FreedomHouse2024IranFOTN,
  author       = {{Freedom House}},
  title        = {Iran: Freedom on the Net 2024 Country Report},
  year         = {2024},
  note         = {\href{https://web.archive.org/web/20260124183945/https://freedomhouse.org/country/iran/freedom-net/2024}{https://freedomhouse.org/country/iran/freedom-net/2024} (Archived on 2026-01-24)},
}

@online{filterwatch2026regional,
  title        = {From Regional Disruptions to Total Blackout: Examining Iran’s Internet During Escalating Protests},
  author       = {{FilterWatch}},
  year         = {2026},
  month        = {January},
  day          = {9th},
  XXXnote         = {Analysis of layered disruptions and escalation to nationwide blockages},
  note         = {\href{https://web.archive.org/web/20260221051736/https://filter.watch/english/2026/01/09/network-monitoring-january-2026-internet-repression-in-times-of-protest/}{https://filter.watch/english/2026/01/09/network-monitoring-january-2026-internet-repression-in-times-of-protest/} (Archived on 2026-02-21)}
}

@techreport{filterwatch2026_breakdown,
  title        = {IRAN: 2026 Shutdown Technical Analysis},
  institution = {{FilterWatch}},
  author       = {{FilterWatch} and {IODA} and {Kentik} and {Miaan Group} and {ASL19}},
  year         = {2026},
  month        = {January},
  day          = {16th},
  note         = {\href{https://web.archive.org/web/20260228055529/https://filter.watch/wp-content/uploads/sites/2/2026/01/Iran-Internet-Shutdown-Technical-Analysis-January-2026.pdf}{https://filter.watch/wp-content/uploads/sites/2/2026/01/Iran-Internet-Shutdown-Technical-Analysis-January-2026.pdf} (Archived on 2026-02-28)}
}

@techreport{filterwatch2026network,
  title        = {A Month of Iran’s Internet: From Regional Disruptions and Blackouts to a New Whitelisted Reality},
  author       = {Narges Keshavarznia},
  institution  = {Filterwatch},
  year         = {2026},
  month        = {January},
  day          = {28},
  type         = {Technical Report},
  urlXXX          = {https://filter.watch/english/2026/01/28/network-monitoring-january-2026-from-regional-disuptions-to-total-blackout-and-whitelisted-access/},
  note = {\href{https://web.archive.org/web/20260223171136/https://filter.watch/english/2026/01/28/network-monitoring-january-2026-from-regional-disuptions-to-total-blackout-and-whitelisted-access/}{https://filter.watch/english/2026/01/28/network-monitoring-january-2026-from-regional-disuptions-to-total-blackout-and-whitelisted-access/} (Archived on: 2026-01-28)}
}

@techreport{afpc2025riseOfNIN,
  title        = {Iran’s Digital Fortress: The Rise of the National Information Network},
  institution  = {American Foreign Policy Council},
  author       = {{Calla O’Neil}},
  year         = {2025},
  month        = {August},
  number       = {16},
urlxxx          = {https://web.archive.org/web/20260228181911/https://www.afpc.org/uploads/documents/Iran_Strategy_Brief_No._16_-_August_2025.pdf},
  note         = {\href{https://web.archive.org/web/20260228181911/https://www.afpc.org/uploads/documents/Iran_Strategy_Brief_No._16_-_August_2025.pdf}{https://www.afpc.org/uploads/documents/Iran\_Strategy\_Brief\_No.\_16\_-\_August\_2025.pdf}},
}

@techreport{filterwatch2026december2025,
  title        = {Connected but Unsafe: The Model of Regional Internet Repression During the December 2025 – January 2026 Protests},
  author       = {{Narges Keshavarznia}},
  institution  = {Filterwatch},
  year         = {2026},
  month        = {January},
  day          = {5},
  type         = {Technical Report},
  url          = {},
  note         = {\href{https://web.archive.org/web/20260305033930/https://filter.watch/english/2026/01/05/network-monitorig-december-2025-internet-repression-in-times-of-protest/}{https://filter.watch/english/2026/01/05/network-monitorig-december-2025-internet-repression-in-times-of-protest/} (Archived on 2026-03-05)},
  accessed     = {2026-02-20}
}

@techreport{filterwatch2025stealth,
  title        = {Iran’s Stealth Blackout: A Multi-Stakeholder Analysis of the June 2025 Internet Shutdown},
  author       = {{Miaan Group, ASL19, IODA}},
  institution  = {},
  year         = {2025},
  month        = {October},
  day          = {2},
  type         = {Technical Report},
  note         = {\href{https://web.archive.org/web/20260222084425/https://filter.watch/wp-content/uploads/sites/2/2026/01/Irans-Stealth-Blackout_-A-Multi-stakeholder-Analysis-of-the-June-2025-Internet-Shutdown.pdf}{\nolinkurl{https://filter.watch/wp-content/uploads/sites/2/2026/01/Irans-Stealth-Blackout_-A-Multi-stakeholder-Analysis-of-the-June-2025-Internet-Shutdown.pdf}} (Archived on 2026-02-22)},
}

@techreport{ioda2022iran,
  title        = {Technical Multi-Stakeholder Report on Internet Shutdowns: The Case of Iran Amid Autumn 2022 Protests},
  author       = {{OONI, IODA, M-Lab, Cloudflare, Kentik, Censored Planet, ISOC, Article 19}},
  institution  = {},
  month        =  {November},
  year         = {2022},
  type         = {Technical Report},
  note         = {\href{https://web.archive.org/web/20260116141328/https://ioda.inetintel.cc.gatech.edu/reports/technical-multi-stakeholder-report-on-internet-shutdowns-the-case-of-iran-amid-autumn-2022-protests/}{\nolinkurl{https://ioda.inetintel.cc.gatech.edu/reports/technical-multi-stakeholder-report-on-internet-shutdowns-the-case-of-iran-amid-autumn-2022-protests/}} (Archived on 2026-01-16 )},
}

\end{document}